\documentclass[%
 reprint,
 amsmath,amssymb,
 aps,superscriptaddress,
]{revtex4-2}

\usepackage{graphicx}
\usepackage{subcaption}
\usepackage{amsmath}
\usepackage{bm}
\usepackage{amsfonts}
\usepackage{amssymb}
\usepackage{hyperref}
\usepackage{verbatim}
\usepackage{float}

\usepackage{xcolor}
 
\usepackage{multirow}
\usepackage[section]{placeins}

\begin{document}

\title{Light Field Ghost Imaging}

\author{A. Paniate}
\altaffiliation{These authors contributed equally to this work.}
 \affiliation{DISAT, Politecnico di Torino, Corso Duca degli Abruzzi 24, 10129 Torino, Italy}
  \affiliation{Quantum metrology and nano technologies division, INRiM, Strada delle Cacce 91, 10135 Torino, Italy}
 
\author{G. Massaro}
\altaffiliation{These authors contributed equally to this work.}
\affiliation{Dipartimento Interateneo di Fisica, Universit\`{a}  degli Studi di Bari Aldo Moro, 70125 Bari, Italy}
\affiliation{Istituto Nazionale di Fisica Nucleare (INFN), Sezione di Bari, 70125 Bari, Italy}

\author{A. Avella}
 \email{a.avella@inrim.it}
  \affiliation{Quantum metrology and nano technologies division, INRiM, Strada delle Cacce 91, 10135 Torino, Italy}

\author{A. Meda }
  \affiliation{Quantum metrology and nano technologies division, INRiM, Strada delle Cacce 91, 10135 Torino, Italy}

\author{F. V. Pepe}

\affiliation{Dipartimento Interateneo di Fisica, Universit\`{a}  degli Studi di Bari Aldo Moro, 70125 Bari, Italy}
\affiliation{Istituto Nazionale di Fisica Nucleare (INFN), Sezione di Bari, 70125 Bari, Italy}

\author{M. Genovese}
  \affiliation{Quantum metrology and nano technologies division, INRiM, Strada delle Cacce 91, 10135 Torino, Italy}

  \author{M. D'Angelo}
\altaffiliation{Equal last author contribution.}
\affiliation{Dipartimento Interateneo di Fisica, Universit\`{a}  degli Studi di Bari Aldo Moro, 70125 Bari, Italy}
\affiliation{Istituto Nazionale di Fisica Nucleare (INFN), Sezione di Bari, 70125 Bari, Italy}

\author{I. Ruo-Berchera}
\altaffiliation{Equal last author contribution.}
  \affiliation{Quantum metrology and nano technologies division, INRiM, Strada delle Cacce 91, 10135 Torino, Italy}

\begin{abstract}
Techniques based on classical and quantum correlations in light beams, such as ghost imaging, allow us to overcome many limitations of conventional imaging and sensing protocols. Despite their advantages, applications of such techniques are often limited in practical scenarios where the position and the longitudinal extension of the target object are unknown. In this work, we propose and experimentally demonstrate a novel imaging technique, named Light Field Ghost Imaging, that exploits light correlations and light field imaging principles to enable going beyond the limitations of ghost imaging in a wide range of applications. Notably, our technique removes the requirement to have prior knowledge of the object distance allowing the possibility of refocusing in post-processing, as well as performing 3D imaging while retaining all the benefits of ghost imaging protocols.
\end{abstract}

\maketitle

\section{Introduction}

Correlations in light beams have been explored in both quantum and classical context to overcome the limitations of conventional optical measurements \cite{Maccone2011, losero2018loss, Pradyumna2020, Zavatta2006, Iskhakov2011special, Clark2021special, avella2016absolute, Agliati2005, allevi2012measurung, bondani2007sub, allevi2012high} and in particular of imaging \cite{pittman1995optical,gatti2004ghost,dangelo2005quantum,brida2011systematic,valencia2005two,scarcelli2006can,genovese2016real,schwartz2013superresolution,israel2017quantum,dertinger2009fast,lemos2014quantum,dangelo2004identifying,scarcelli2007random,bennink2002two,thiel2007,ono2013,Agafonov2009,meda2015magneto, TammaOE,TammaSR}. In the quantum domain, correlation and entanglement have been demonstrated to improve the sensitivity in imaging of amplitude and phase samples \cite{moreau2019imaging} enabling real-time sub-shot-noise microscopy \cite{samantaray2017realization} and pattern recognition \cite{ortolano2023}. However, correlations properties relevant to imaging can also be found in specific kinds of classical beams, and many protocols originally developed in the quantum domain have been shown to work regardless of the origin, either quantum or classical, of the correlation \cite{gatti2004ghost,valencia2005two,bennink2002two,thiel2007}. Still, the imaging performances enabled by classical correlations tend to be outperformed by quantum ones, especially in the low photon number regime \cite{losero2019differential}; a relevant example is the impossibility of achieving sub-shot-noise sensitivity by means of classical correlations \cite{brida2011imaging, GattoMonticone2014, brida2010experimental, lopaeva2013}.

One of the most celebrated techniques that came out in this context is ghost imaging (GI). In GI \cite{pittman1995optical,gatti2004ghost,dangelo2005quantum,valencia2005two, scarcelli2006can,brida2011systematic}, two correlated beams are used: one beam propagates towards a ``reference'' spatially-resolving detector, either freely or through optical elements; the other beam illuminates the object of interest, and a ``bucket'' detector collects a signal proportional to the \textit{total} intensity of light transmitted, reflected, scattered, or even transduced by the object. The image is not directly formed on the spatially resolving sensor, but is instead reconstructed by correlating the fluctuations of intensities registered by the reference and the bucket detectors. This is possible since the spatial intensity pattern on the object is correlated in space and time with the one impinging on the reference detector. In such a scheme, the bucket detector is not meant to detect the accurate spatial distribution of light on the object, hence, no cumbersome optical imaging systems and spatial resolving detectors are needed in the object arm; on the other hand, the propagation of the correlated beam toward the spatially-resolving detector can occur through a different optical path in a controlled environment. For this reason, GI is much less sensitive than conventional imaging to detrimental effects connected to the propagation from the object to the sensor. This property makes GI-based protocols particularly interesting when either propagation from the object is heavily disturbed, or the object itself converts the impinging electromagnetic field into a signal of a different nature (e.g., a neuronal pulse). However, a drawback of GI is the need to acquire a large number of frames to compute the correlations; this makes GI relatively slow compared to imaging techniques based on direct intensity detection. In particular, when the object distance is not known, as one may expect in real applications, it is not possible to adapt focusing in real-time: many blurred ghost imaging would need to be acquired while tentatively changing the focus plane of the imaging system, till the correct conjugate plane is identified. Of course, increasing the native depth of field is always possible, but this goes at the expense of giving up longitudinal resolution, hence, 3D reconstructions. Arguably, this is one of the main reasons why GI and its variants have not yet overcome the gap to widespread application.

In traditional (not quantum) optics, a direct technique called plenoptic imaging or light field imaging (LFI) enables the user to reconstruct (or refocus) out-of-focus parts within the acquired image and to change the point of view, in post-processing ~\cite{lippmann1908epreuves,adelson1992single,ng2005light}. This is enabled by the simultaneous detection of the spatial distribution and the propagation direction of light, which is achieved by placing a micro-lens array (MLA) between the main lens and the sensor of a standard imaging device~\cite{georgiev2009high,georgiev2010focused,georgiev2012multifocus,
goldlucke2015plenoptic,jin2017point, ng2005light,ng2005fourier}. The MLA forms sub-images by rays coming from different portions of the main lens, thus offering different perspectives on the scene of interest. As a drawback, the spatial resolution is decreased with respect to the diffraction limit, as defined by the numerical aperture of the main lens; such resolution loss is, in fact, proportional to the number of directional resolution cells. Still, PI is one of the simplest and fastest state-of-the-art methods to achieve volumetric images~\cite{levoy2006light,levoy2009recording,cheng2011simultaneous,
abrahamsson2012fast,quirin2013instantaneous,broxton2013wave,xiao2013advances,prevedel2014simultaneous,ren2017fast,dansereau2013decoding,adhikarla2015exploring,wanner2012globally}. Hence, 
despite the aforementioned limitation, PI is increasingly employed in the diversified tasks, including 3D imaging and sensing~\cite{liu20153d,xiao2013advances}, stereoscopy~\cite{adelson1992single,muenzel2013enhancing,levoy1996light}, particle image velocimetry~\cite{fahringer2015volumetric}, particle tracking and sizing~\cite{hall2016comparison}, wavefront sensing~\cite{prevedel2014simultaneous,lv2016g,wu2016using,wu2016imaging},  and microscopy~\cite{levoy2006light,glastre2013demonstration,
prevedel2014simultaneous,broxton2013wave}. 
Recently, an interesting correlation technique, called \textit{correlation plenoptic imaging} (CPI), has been developed, with the specific purpose of improving volumetric resolution with respect to direct PI methods \cite{dangelo2016correlation,pepe2016correlation,pepe2017diffraction,dilena2018correlation,scagliola2020correlation,massaro2021lightfield,dilena2020correlation,abbattista2021towards,massaro2022comparative,massaro2022effect,scattarella2022resolution,massaro2022refocusing,massaro2023correlated}. CPI exploits the simultaneous  momentum and position correlation in two classical or quantum beams to obtain, at the same time, high spatial and directional resolution; this is achieved by correlating light intensities measured by two disjoint detectors, each one imaging a different plane, at a different optical distance from the source. 

However, both LFI and CPI require that a spatially resolving detector is placed in the object's optical path, and that light propagating from the scene to the sensors behaves in a predictable way, as scattering and distortion effects would hinder the one-to-one correspondence between image and object point. 

In this article, we show that the principles of LFI and GI can be at the basis of a new technique called \textit{Light Field Ghost Imaging} (LFGI), which combines the use of a bucket-detector, in the object arm, with the availability of directional information provided by the insertion of a MLA before the spatially-resolving sensor. Such a realization represents the first proper adaptation of plenoptic imaging to GI tasks. LFGI shares the benefits of both GI and LFI: its refocusing capabilities enables to overcome the impossibility of standard GI for real-time focusing, and its bucket-detection offers increasing robustness to distortion, scattering and noise in the surrounding of the object, with respect to direct PI. Moreover, while extending the depth of field of GI, LFGI preserves the longitudinal resolution, thus enabling imaging and ranging capabilities, as required in diverse applications, from remote sensing to microscopy.  

Here, we demonstrate LFGI with thermal correlated beams, but, with the appropriate differences in the optical schemes, LFGI can be used with any kind of correlated light beams, including the ones formed by entangled photons generated by spontaneous parametric down-conversion (SPDC). In the last case, a significant advantage in terms of signal-to-noise could be eventually achieved in the low illumination regime, as in the case of GI \cite{meda2017photon}. Still, our choice of employing thermal light is meant to overcome the limitations connected with entangled photon generation and detection, i.e., the low production rate, and connected larger measurement time (see Refs.~\cite{massaro2023correlated,defienne2021fullfield} for a comparison of imaging protocols employing ultrafast detectors).

\section{Results}

\begin{figure*}
	\centering
	\includegraphics[scale=0.6]{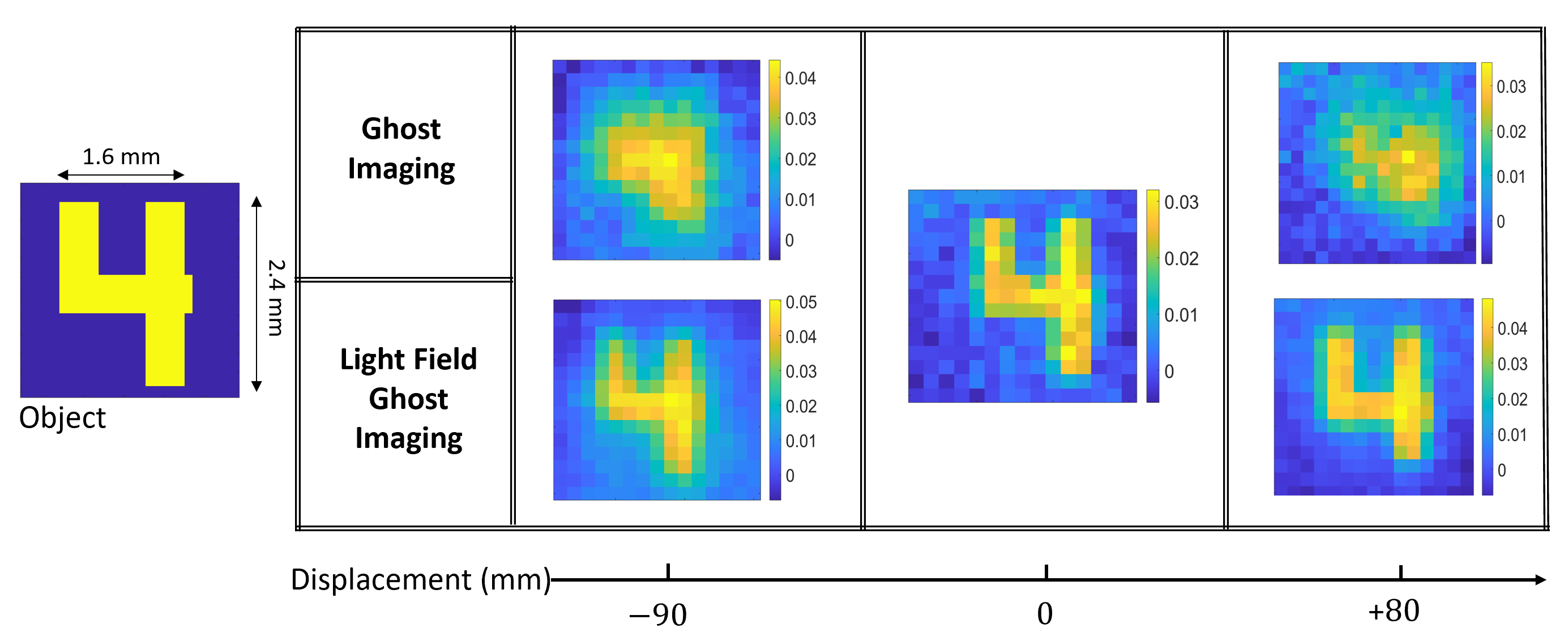}
	\caption{Experimental reconstructions of the image of a ``4"-shaped two level mask as a function of the displacement between the sample and the GI on focus plane. For each displacement $5 \cdot 10^4$ independent patterns are acquired with integration time of 15 ms.} 
 \label{ResolutionGraphFour}
\end{figure*}

A qualitative example of the refocusing capability of LFGI is shown in Fig.\ref{ResolutionGraphFour}, where an experimental image of a ``4"-shaped sample reconstructed with both LFGI and GI is reported. In this exemplifying result, the sample is placed at different longitudinal distances from the object plane that would give a focused GI, which shall indicate as "GI on focus plane".

When the sample is placed in the GI on focus plane, i.e. without displacement, GI retrieves the object correctly, and there is no difference between GI and LFGI.
On the contrary, for large displacements, GI fails to produce a sharp image: the acquired blurred images loose most of the information about the object. This effect is caused by the loss of correlation between the detected and the probing intensity pattern on the object, due to the different propagation distance. Conversely, thanks to its refocusing ability, LFGI enables to identify the corresponding pattern that probes the object and thus to retrieve the object details accurately.

\begin{figure*}
	\includegraphics[scale=0.55]{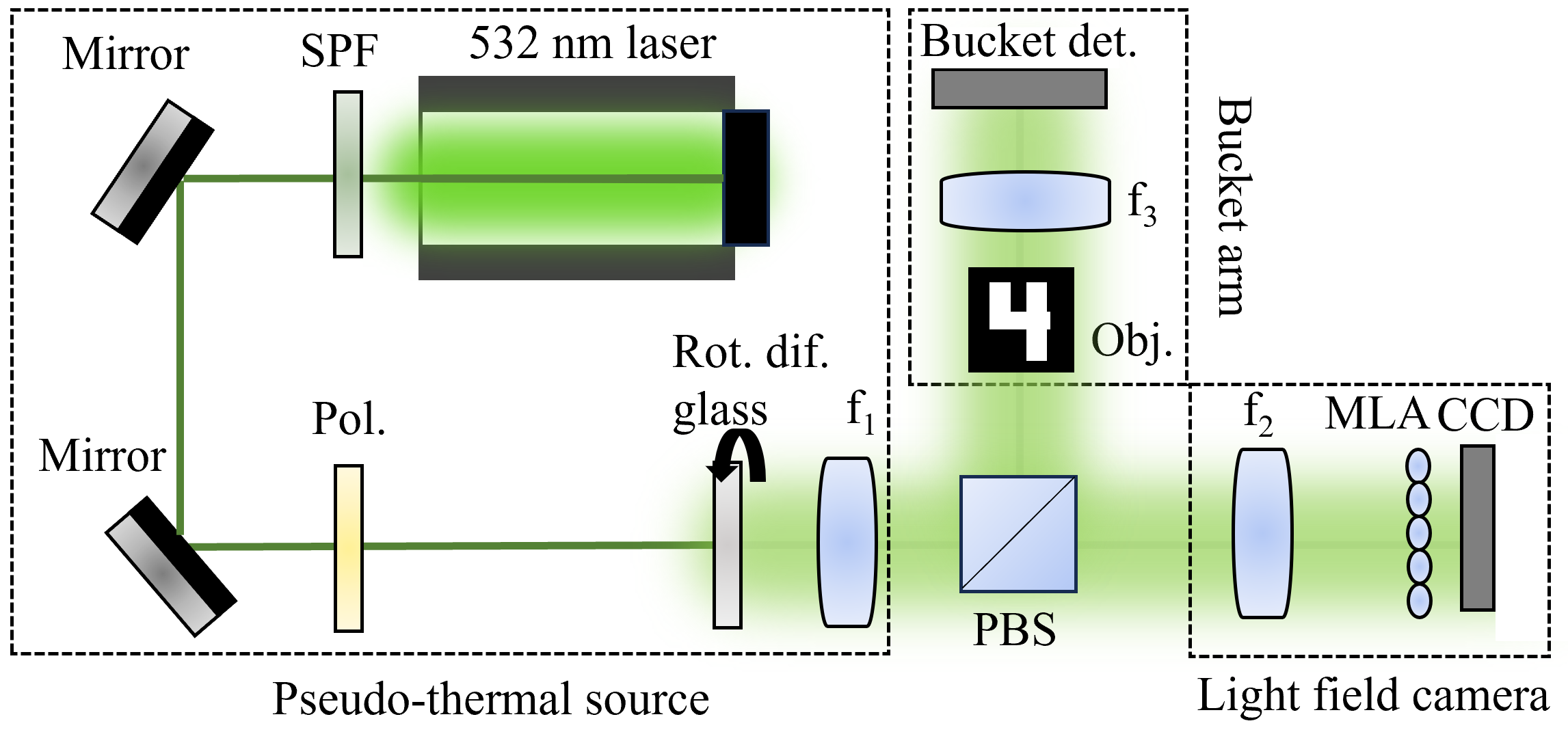}
	\caption{
A collimated beam from a CW laser at 532 nm, after being spatially (SPF) and polarization filtered (Pol.), is shined onto a rotating diffusing glass (Arecchi's disk). The scattered light is collected by a far-field lens with focal length $f_1 = 75 $ mm. The disk and the far-field lens form random speckle patterns at the polarizing beam splitter (PBS). The rotation of the glass generates, in time, different patterns that are equally split by the PBS with a polarization angle of $\pm 45^{\circ}$ with respect to the axis of the linear polarizer. 
One beam is imaged by a light field camera, composed of a lens with focal length $f_2 = 80$ mm and a micro-lens array (MLA) of dimension 30 $\times$ 30. Each micro lens has a focal length $f_{MLA} = 14.6$ mm and a diameter $d=300 \ \mu$m. The MLA is placed at a distance equal to its focal length $f_{MLA}$ from a charge-coupled-device (CCD) camera (Andor Luca R (604)). 
The other beam probes the object and is imaged with a lens of focal length $f_3 = 75 $ mm to a CCD camera (Andor iXon Ultra): the bucket detector is obtained by integrating the signal from all illuminated pixels. 
}
    \label{scheme}
\end{figure*}

The results shown in Fig. \ref{ResolutionGraphFour} are obtained by employing the experimental scheme depicted in Fig. \ref{scheme}. In this setup, we split the speckled beam generated by a so-called pseudo-thermal source to obtain two classically correlated intensity patterns, with a characteristic spatial speckled structure \cite{gatti2008three}. The speckle pattern has a diameter of $2$ cm and each speckle has an averaged transversal size $\sim 8 \ \mu$m and an averaged longitudinal length $\sim 1 \ $mm.
One of the correlated beams is sent to the bucket arm. Here, the beam probes the sample and the transmitted light is collected by a bucket detector which provides a signal proportional to the intensity of the total incident light. The other beam is sent to a standard light field camera \cite{hahne2014light} which is essentially composed of a lens, a MLA and a spatial resolving detector (CCD).
The plenoptic camera acquires the combination of the angular and spatial information that is necessary for the 3D reconstruction of the GI signal; to this end, each pixel of the reference arm is correlated with the integrated signal from the bucket detector, as rigorously described in the Methods section. In practice, intuitively, the process may be represented as a two-step algorithm. In the first step, the complete space-momentum information at the PI camera is used to reconstruct the patterns of speckles along the propagation axis at different longitudinal distances. In the second step, one correlates the reconstructed speckle patterns with the bucket detector, as done in standard GI; this enables retrieving a 3D reconstruction of the objects in the bucket arm without knowing, a priori, their positions along the propagation axis. 

\begin{figure*}
	\centering
	\includegraphics[scale=0.54]{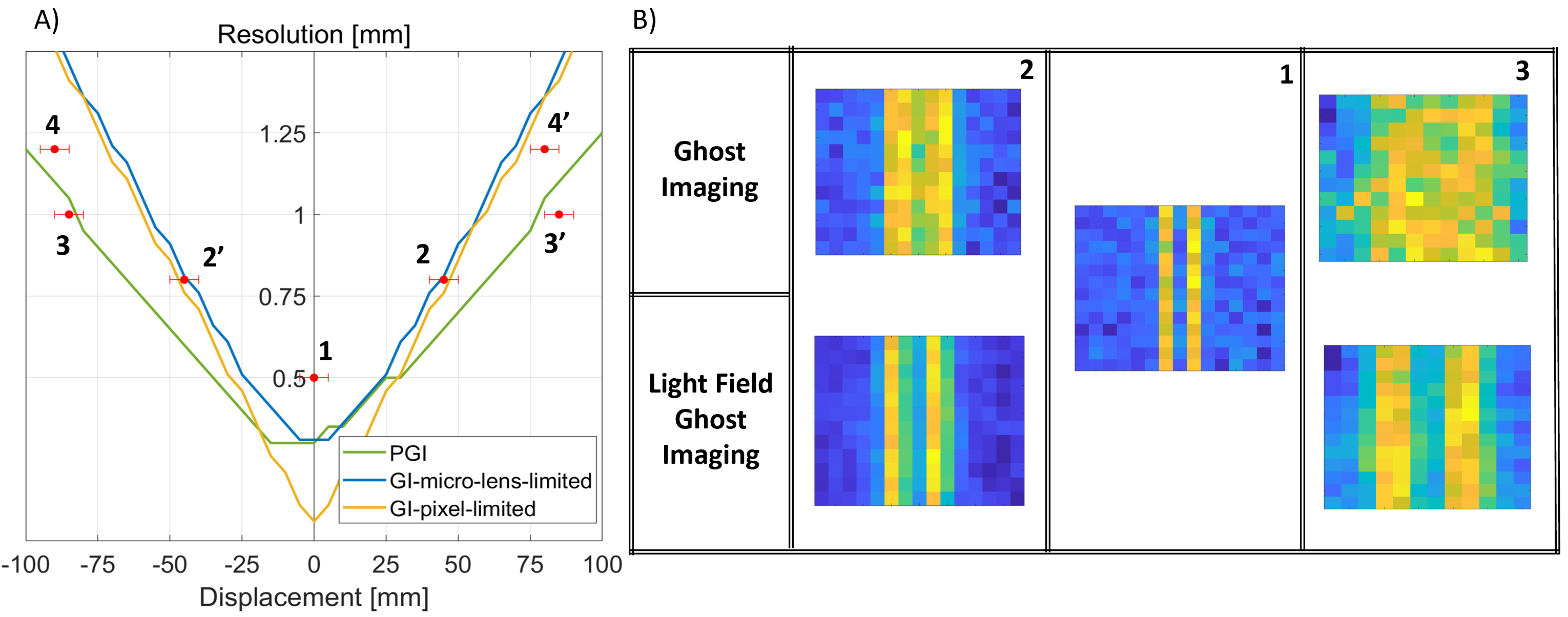}
	\caption{\textit{Resolution of GI and LFGI as a function of the displacement.} \textbf{A} The curves represent the set of points where the visibility of a double-slit is exactly 40\%, for the GI-micro-lens-limited, the GI-pixel-limited, and the LFGI reconstruction, respectively. \textbf{B} Examples of GI and LFGI reconstructions of double-slit masks with different separations and different displacements, as indicated by the red circles in (A). The same conclusions can be derived from the other points (`` 2' ",`` 3' ",`` 4 ", `` 4' ") that are not reported for shortness.} All LFGI reconstructions show visibility greater or equal to $40 \%$ \label{ResolutionGraph}
\end{figure*}
A more systematic and quantitative analysis of the quality of the LFGI reconstructions versus standard GI is reported in Fig.\ref{ResolutionGraph}A, showing the resolution achievable as a function of the displacement. The resolution curves are calculated as the minimum distance between the centers of two slits that gives rise to a resolved image with visibility of $40 \%$. 
For a comprehensive analysis, LFGI is compared with both the GI characterized by the same minimal resolution of the plenoptic camera, which is
set by the dimension of the micro-lenses (GI-micro-lens-limited), and the GI with resolution limited only by the pixel size (GI-pixel-limited), as it would be without the MLA. The three curves represent, respectively, the theoretical minimal resolution achievable with LFGI (green line), the GI-micro-lens-limited (blue line) and the GI-pixel-limited (yellow line). Note that, while the GI-pixel-limited resolution is better nearby the zero-displacement plane, it approaches the GI-micro-lens-limited case for larger displacement and becomes worse than in LFGI already for displacement as small as $20$ mm.  Experimentally, all the reported GI reconstructions are retrieved as GI-micro-lens-limited in order to avoid changes of the experimental setup that can affect the final results. \\

\begin{figure*}
	\centering
	\includegraphics[scale=0.6]{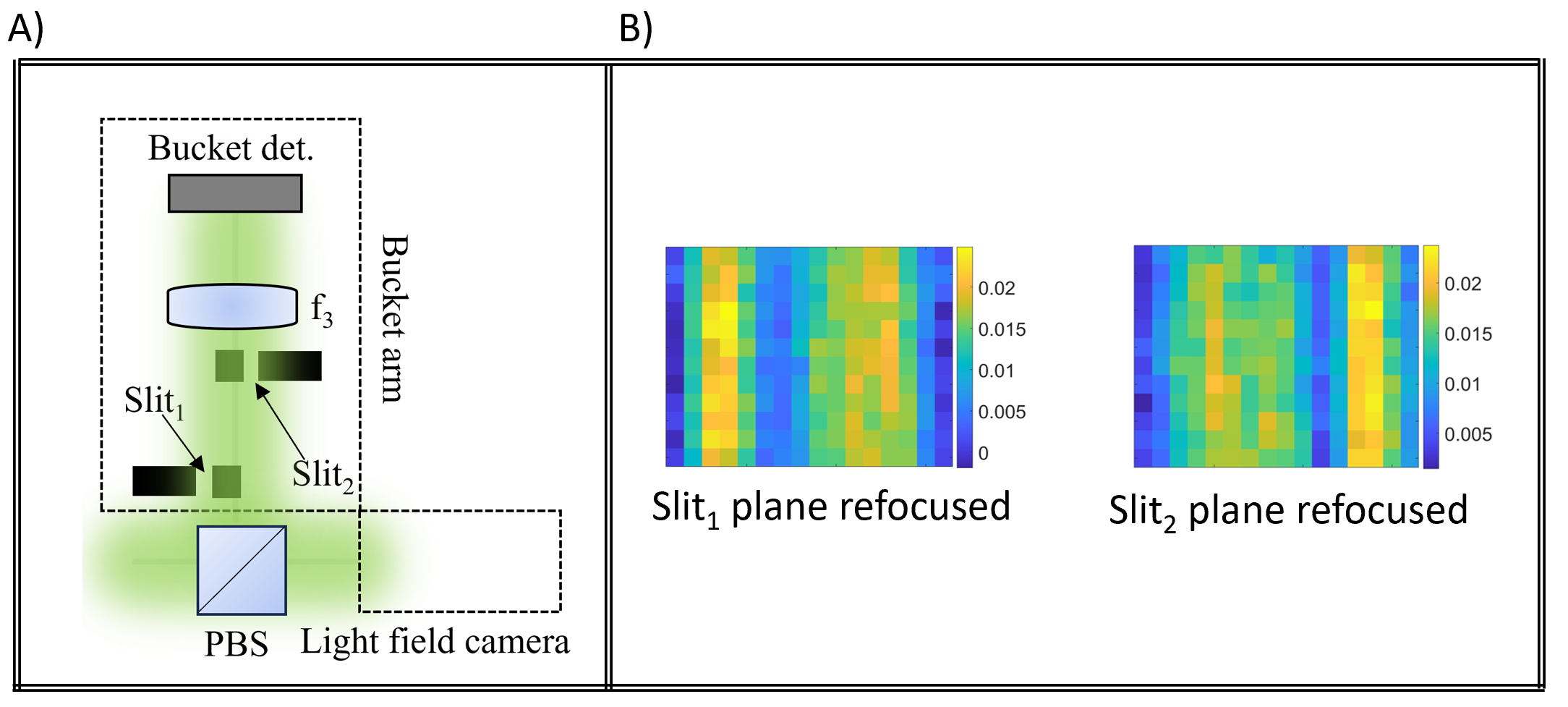}
	\caption{\textit{Refocusing ability of LFGI.} \textbf{A} Two 1 mm-width slits ($\text{slit}_1$ and $\text{slit}_2$) are placed with displacement -80 mm and +80 mm, respectively, in the configuration shown in the figure. \textbf{B} In the picture on the left, LFGI is employed to refocus $\text{slit}_1$ while $\text{slit}_2$ is out of focus; conversely, in the right-hand side, LFGI refocuses the plane of $\text{slit}_2$. }\label{Refocus}
\end{figure*}

Different double-slit masks (from NBS 1963A Thorlabs) at different displacements have been used in order to experimentally verify the LFGI advantage. The coordinates of the dots in Fig. \ref{ResolutionGraph}A represent the displacement and slit separation (the distance between the centers of the two slits) used, respectively.  For example, the dot label as ``1" represents a double-slit with separation of $0.5$ mm, placed in the GI on focus plane, i.e. at zero-displacement. The reconstruction of the object ``1" is shown in the central panel of Fig.\ref{ResolutionGraph}B. In this case, the LFGI and GI-micro-lens-limited reconstructions coincide, and visibility is well above 40\%. For the case represented by the dot ``2" in panel A, the GI and LFGI images are shown in the left-hand side panel of Fig.\ref{ResolutionGraph}B. The image is well reconstructed by LFGI while the visibility is quite poor for GI, in agreement with the theoretical prediction. In the right-hand side panel, we show the results obtained when the displacement and the slit-separation are chosen to be on the limit of the 40\% visibility curve of LFGI (green line), as represented by the points ``3" in panel A, and well below the corresponding GI visibility (yellow and blue lines): The GI reconstruction is completely blurred, while the slits appear well resolved in LFGI.
The plenoptic advantage of LFGI over the GI micro-lens-limit is confirmed in terms of resolution for every displacement; with respect to the GI-pixel-limited case, the advantage appears for displacements $\gtrsim 20$ mm due to the loss of resolution introduced by MLA of the LFI camera.
Moreover, the enhancement provided by LFGI over GI is not limited to the mere extension of the depth of field (DOF) of GI, as shown in Fig.\ref{ResolutionGraph}A. In fact, the unique feature of PI is the possibility to refocus on different planes, after the acquisition, without loosing longitudinal resolution. LFGI thus provides the depth information of the reconstructed objects, as demonstrated in Fig. \ref{Refocus}.  

Two single slits, $\text{slit}_1$ and  $\text{slit}_2$, are placed at different distances from the PBS, at +80mm and -80mm, respectively. The 3D reconstruction of LFGI allows refocusing a-posteriori on both sets of planes corresponding to $\text{slit}_1$ and to $\text{slit}_2$. In Fig.\ref{Refocus}B, the LFGI reconstructions of the two single slits is reported: In the left picture, we refocused on $\text{slit}_1$ and, as a result, $\text{slit}_2$ results out-of-focus; in the right picture, we refocused on $\text{slit}_2$ and $\text{slit}_1$ appears completely blurred. This example shows the sectioning capability of a 3D scene enabled, in post-processing, by LFGI.

\section{Methods}

The information on the reconstructed images is encoded in the plenoptic function
\begin{equation}\label{eq:plenopticART}
    P(\bm{x})=\left\langle \left( I_{B} - \left\langle I_{B} \right\rangle \right) \left( I(\bm{x}) -  \left\langle I(\bm{x}) \right\rangle \right) \right\rangle ,
\end{equation}
obtained by correlating the fluctuations of the intensity $I_B$ acquired by the bucket detector, which is proportional to the total intensity of light that propagates from the object, with the fluctuations of the intensity registered by the pixel centered on the transverse coordinate $\bm{x}$ of the reference sensor. For mitigating the effect of the noisy background typical of thermal light GI, we shall actually replace $I_B$ with a ``differential'' signal \cite{ferri2010differential, losero2019differential}.

\begin{figure*}[th]
	\centering
    \includegraphics[ width=1\textwidth]{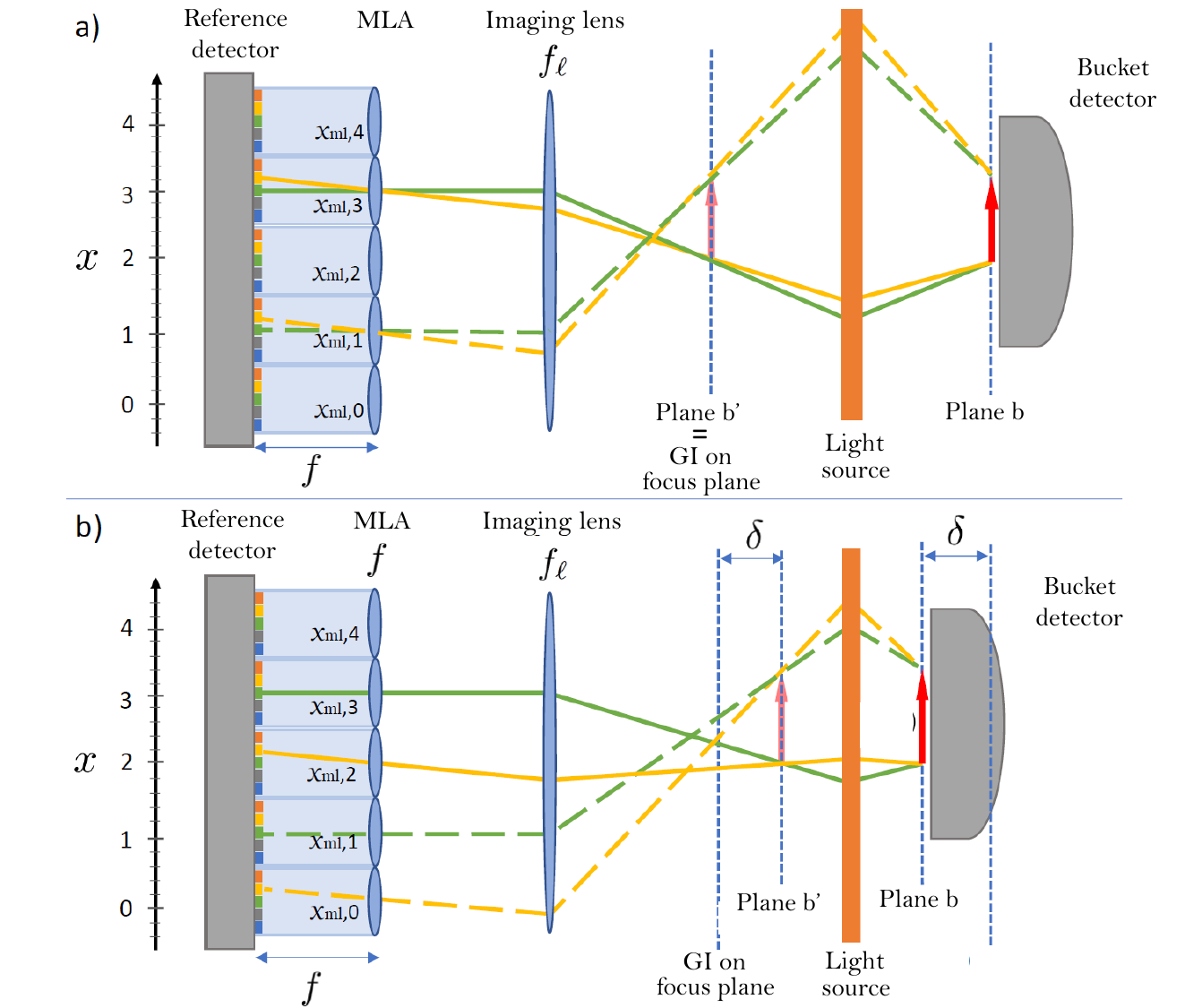}
	\caption{Schematic representation of the principle of LFGI. Panel a) shows the scheme when the object is imaged (through correlation measurements) in the zero-plane of the light field camera (i.e., the GI is at focus). Figure b) shows the scheme when the object is displaced by a distance $\delta$ with respect to the case a). 
    The difference between the two cases gives rise to different optical paths for the two pairs of rays emitted by the extreme points of the object. In the focused case, rays emitted by the same object point arrive on the same microlens, regardless of the emission angle. Plenoptic information is irrelevant in this situation.
    On the contrary, when the ghost image is out of focus, different microlenses are crossed, and specific pixels are illuminated depending on the angle of emission.}\label{principlescheme}
\end{figure*}

The conceptual scheme of Fig. \ref{principlescheme} provides an intuitive picture of the working principle of LFGI. Here, for convenience, the source of correlated light beams is compressed in the vertical thick orange bar and the two correlated beams are shown, respectively, on the right and on the left of the source, along the line of Klyshko picture \cite{Klyshko, PhysRevLett.94.063601, Aspden2014Klyshko}. 
In this picture, we call \textit{plane b} the plane of the object in the bucket arm and \textit{plane $b^\prime$} the plane correlated with the object plane in the reference arm. 
Let us start by considering, in the geometric optics approximation, two explanatory cases. In the first case (Fig. \ref{principlescheme} (a), \textit{plane $b^\prime$} correspond to the GI on focus plane (i.e. the plane whose image, realised by the imaging lens of the light field camera, is in focus on the plane of the micro-lenses). Consequently, all rays correlated with a particular point of the object fall on the same micro-lens. Therefore, in Eq. (\ref{eq:plenopticART}), we can evaluate $I(x)$ by integrating all pixels below each micro-lens. On the contrary, when there is a displacement $\delta$ between the \textit{plane $b^\prime$} and the GI on focus plane, as in Fig. \ref{principlescheme} (b), rays correlated with the same point of the object fall on different micro-lens. However, the coordinate of the pixels below each micro-lens identify precisely the light propagation direction, which we represent by a different color in the picture \cite{hahne2016refocusing}; this allows to obtain information on the whole 3D light-field propagating from the source. The algorithm to obtain $I(x)$ in the more interesting case (b) where displacement is non-zero shall now be identified through an accurate analysis of the LFGI scheme.

Due to the statistical properties of the source, the correlation between the intensity fluctuations of Eq. \eqref{eq:plenopticART} reduces to
\begin{equation}
    P(\bm{x})\simeq\int 
        \left\lvert
            \left\langle
                E^*_\text{ref}(\bm x)
                E_B(\bm x_b)
            \right\rangle
        \right\rvert^2
        d\bm x_b,
        \label{eq:plenoptic}
\end{equation}
\noindent where $E^*_\text{ref}$ and $ E_B$ are the electric fields at the reference and bucket detector, respectively. Without loss of generality, we can assume to put the bucket detector immediately behind the object, so that  $ E_B$ is also the electric field at the object. 
$E_\text{ref}$ and $ E_B$ can be obtained by propagating  respectively the field at the \textit{plane $b^\prime$} and \textit{$b$} through the two optical paths, which gives:
\begin{multline}\label{Eref}
    E_\text{ref}(\bm x) = \\  
        = \int  E(\bm x_{b}^\prime)\,
        e^{ik\frac{(\bm x_o - \bm x_b^\prime)^2}{2\delta}}\,
        \mathcal{P}\left(\bm x_o+\frac{\bm x_i}{M}\right) \mathcal{L}(\bm x,\bm x_i) 
        d\bm x_o d\bm x_i d\bm x_b^\prime 
\hspace*{-10pt}
\raisetag{35pt}
\end{multline}

\begin{equation}
 \label{Eb}
    E_B(\bm x_b) = \mathcal{A}(\bm x_b)\cdot
    E(\bm x_b)
\end{equation}
\noindent where $E(x_b^\prime)$ is the speckled
electric field at the plane $b^\prime$, $e^{ik\frac{(\bm x_o - \bm x_b^\prime)^2}{2\delta}}$ is the free propagation between the \textit{plane $b^\prime$} and the GI on focus plane, 
$\mathcal P(\bm x)$ is the point-spread function of the imaging system from the GI on focus plane to a plane immediately before the MLA, $M$ is its magnification, and $\mathcal{L}(\bm x,\bm x_i)$ is the propagation function from the MLA to the detector plane; this function is a linear superposition of the propagation functions $\mathcal L_{\bm x_\text{ml}}$ associated with each micro-lens (see Fig. \ref{principlescheme}):
 \begin{equation}
\mathcal{L}(\bm x,\bm x_i)= \sum_{\bm x_{ml}}\mathcal L_{\bm x_\text{ml}}(\bm x,\bm x_i)
=\sum_{\bm x_{ml}}f_{\bm x_{ml}}(\bm x_i)\, 
        e^{ik\frac{(\bm x - \bm x_i)^2}{2f}}
\end{equation}
with $f$ the micro-lens focal length, supposed equal throughout the whole array, and
\begin{equation}
 f_{\bm x_{ml}}(\bm x_i) = 
    \begin{cases}
        e^{-ik\frac{(\bm x_i - \bm x_{ml})^2}{2f}}&\text{if }\lvert\bm x_i - \bm x_{ml}\rvert^2\le (\Delta x_{ml})^2\\
        0 & \text{elsewhere}
    \end{cases}
\end{equation}
is the trasmission function of each microlens centered around $\bm x_{ml}$ with pitch $2\Delta x_{ml}$.
 In Eq. (\ref{Eb}), $\mathcal A(\bm x_b)$ is the object field transmittance, whose square module is reconstructed through refocusing, and $E(\bm x_b)$ is the speckled electric field impinging on the object. 
The analytical expression of the plenoptic function is derived by plugging the expressions of $E_\text{ref}$ and $E_B$ into Eq. (\ref{eq:plenoptic}).

In order to actually obtain plenoptic information from $P(\bm x)$, which is currently a 2D function, one must ensure there is no cross-talk between field intensities from different micro-lenses on the reference sensor. To account for this condition, we decompose $E_\text{ref}$ in terms of the contributions from each micro-lens
\begin{equation}
    E_\text{ref}(\bm x) = \sum_{\bm x_{ml}} E_{\bm x_{ml}}(\bm x),
\end{equation}
where
\begin{multline}
    E_{\bm x_{ml}}(\bm x)= \\
    = \int 
        E(\bm x_b^\prime)\,
        e^{ik\frac{(\bm x_o - \bm x_b^\prime)^2}{2\delta}}\,
        \mathcal{P}\left(\bm x_o+\frac{\bm x_i}{M}\right) \mathcal{L}_{\bm x_{ml}}(\bm x,\bm x_i) 
        d\bm x_o d\bm x_i d\bm x_b^\prime.
\hspace*{-22pt}
\raisetag{35pt}
\end{multline}
When there is no intensity cross-talk, i.e., when $\int \lvert E_{\bm x_{ml}^\prime}(\bm x)\rvert^2\rvert E_{\bm x_{ml}}(\bm x)\lvert^2 d\bm x\simeq 0$ for all $\bm x_{ml}^\prime\neq\bm x_{ml}$, separate non-zero regions can be recognized in $P(\bm x)$, one for each microlens, with analytical expressions
\begin{equation}
    P_{\bm{x}_{ml}}(\bm{x})\simeq \int\left\lvert
            \left\langle
                E_{\bm{x}_{ml}}^*(\bm x)
                E_B(\bm x_b)
            \right\rangle
        \right\rvert^2
        d\bm x_b,
\end{equation}
 Hence, the microlens center, $\bm x_{ml}$, introduces a second variable $\bm x_{ml}$ on which the plenoptic function depends, in the assumption of no cross-talk; we can thus conveniently adopt the redundant notation $P_{\bm{x}_{ml}}(\bm{x})$ to indicate the function in Eq. (\ref{eq:plenopticART}), with $\bm{x}_{ml}$ representing the center coordinate of the microlens whose image is formed on the pixel in position $\bm{x}$. In this way, the plenoptic function becomes explicitly dependent on the expected \textit{four} coordinates.

The plenoptic correlation function can now conveniently expressed in terms of the two-point correlation function $\left\langle E^*(\bm x_b)E(\bm x_b^\prime)\right\rangle=\mathcal S(\bm x_b,\bm x_b^\prime)$, which describes the correlation properties of the speckle patterns at the object ($\bm x_b$) and ghost image ($\bm x_b^\prime$) planes, and depends on the experimental properties of the pseudo-chaotic source and on the object-to-source distance; the result is as follows:

\begin{equation}\label{eq:encodingART}
    P_{\bm{x}_{ml}}(\bm{x}) = \int \Biggl\lvert \int_{S_{ml}}\mathcal{G}(\bm x_b,\bm x_i)e^{-ik\,\bm{x}_i\cdot\frac{\bm{x}-\bm{x}_{ml}}{f}} \,d\bm{x}_{i}\Biggr\rvert^{2}\,d\bm{x}_{b} ,
\end{equation}
where 
\begin{multline}\label{eq:greens}
    \mathcal{G}(\bm x_b,\bm x_i) = \\ = \int\mathcal{A}^*(\bm{x}_{b})\,\mathcal{S}(\bm{x}_{b},\bm{x}_{b}^{\prime})\,e^{ik\frac{(\bm{x}_{o}-\bm{x}_{b}^{\prime})^{2}}{2\delta}}\mathcal{P}\left(\bm{x}_{o}+\frac{\bm{x}_{i}}{M}\right)d\bm{x}_{b}^{\prime}d\bm{x}_{i}
\end{multline}
is the Green's function of conventional (unfocused) GI, and $\delta$ is the defocusing distance (displacement from the plane o). In fact, the standard ghost image $A_\text{GI}(\bm x_i)$ is obtained as \begin{equation} 
    A_\text{GI}(\bm x_i)=\int \left\lvert \mathcal{G}(\bm x_b,\bm x_i)\right\rvert^2d\bm{x}_{b}.
    \label{eq:ghostImage}
\end{equation}
The difference between LFGI and conventional GI is easily understood by comparison of Eqs. (\ref{eq:ghostImage}) and (\ref{eq:encodingART}): whereas the latter is only sensitive to the square module of a given object-dependent quantity $\mathcal G$, the former is also sensitive to the modal (directional) content of $\mathcal G$; this is due to the fact that the integration over the spatial coordinate $\bm x_i$ entails a Fourier transformation in the modal coordinate $k(\bm x - \bm x_{ml})/f$ on the sensor plane.

The image of the object, for each defocusing parameter $\delta$, can be expressed by a function $R_{\delta}$ that is obtained from the plenoptic correlation function  $P_{\bm{x}_{ml}}$ after the definition of the geometrical correspondence between the coordinates of the object and the coordinates in the the reference detector.

According to Eqs.~\eqref{eq:encodingART} and \eqref{eq:greens}, an object point at coordinate $\bm{x}_{b}$ is mapped on the reference detector in the pairs of coordinates $(\bm{x}_{ml},\bm{x})$, satisfying the linear relation
\begin{equation}
    \bm{x}_{b} = \alpha(\delta)\bm{x}_{ml}+\beta(\delta)\bm{x}.\label{eq:linear}
\end{equation}
The coefficients $\alpha(\delta)$ and $\beta(\delta)$  are fixed by the defocusing parameter $\delta$. Their full expression will be derived in the next paragraphs; in the particular case corresponding to $\delta = 0$ (focused case) they reduce to $\alpha(0)=-1/M$ and $\beta(0)=0$. 

As in conventional light field imaging, the operation of refocusing requires collecting the signals from all the points $(\bm{x}_{ml},\bm{x})$ corresponding (within the approximation entailed by the sensor and microlens array granularity) to the same object point $\bm{x}_0$,
\begin{equation}\label{eq:refocusingART}
    R_{\delta}(\bm{x}_{b})= \sum_{\bm x} P_{\frac 1 \alpha \left(\bm{x}_{b}-\beta \bm{x}\right)} \left( \bm x \right),
\end{equation}
where $\alpha$ and $\beta$ are the same $\delta$-dependent coefficients that define the geometrical correspondence of Eq.~\eqref{eq:linear}. In Eq.\eqref{eq:refocusingART}, sub-images $P_{\bm x_{ml}}$ corresponding to a fixed modal coordinate $\bm x$ are shifted and superimposed, so that each pixel in the final image corresponds to a single coordinate of the object plane.\\
The correspondence between object points and microlens-angle pairs is obtained through ray-tracing. The analysis is easily be carried over in a Klyshko picture, tracing rays emitted from the object plane backwards to the source plane, and then forward to the microlenses and reference sensor (see Fig. \ref{principlescheme}). ABCD matrices are a convenient and compact way to express ray tracing, and shall now be used to obtain the final position $x_k$ and angle $\theta^\prime$ on the sensor. To this end we shall indicate with $M_{\text{lens}}(f)$, $M_{\text{vac}}(z)$, $M_{\text{img}}(o,M)$ the matrices for the propagation of the ray, respectively, through a lens of focal distance $f$, a distance $z$ in vacuum, and an imaging system with object distance $o$ and magnification $M$, and further introduce

\begin{equation}
M_{\text{src}}=\left[\begin{array}{cc}
1 & 0\\
0 & -1
\end{array}\right]
\end{equation}
to describe rays from the source, in the Klyshko picture. If a ray is emitted by the object at position $x_{b}$ with angle $\theta$ and passes through the micro-lens centered on $x_{ml}$, its arrival coordinates on the sensor are given by

\begin{multline}
\left[\begin{array}{c}
x_{k}\\
\theta^{\prime}
\end{array}\right]=\left[\begin{array}{c}
x_{ml}\\
0
\end{array}\right]+
M_{\text{vac}}\left(f\right)\cdot M_{\text{lens}}(f)\cdot  \\
\left(M_{\text{img}}(o,M)\cdot M_{\text{src}}\cdot M_{\text{vac}}(\delta)\left[\begin{array}{c}
x_{b}\\
\theta
\end{array}\right]-\left[\begin{array}{c}
x_{ml}\\
0
\end{array}\right]\right).
\end{multline}
Therefore, by expressing both $x_k(x_b,\theta)$ and $\theta^{\prime}(x_b, \theta)$ in terms of the plenoptic coordinates $x$ and $x_{ml}$, a point-to-point correspondence can be obtained between the emission coordinates $(x_k,\theta)$ and the coordinates $(x_{ml}, x)$. One way to relate the arrival coordinates and the plenoptic coordinates is
\begin{equation}
\begin{cases}
x_{k}(x_{b},\theta) & =x\\
\theta^{\prime}(x_{b},\theta) & =(x-x_{ml})/f,
\end{cases}\label{eq:ray_tracing}
\end{equation}
which can be inverted to recover the object coordinate as a function of the plenoptic coordinates $x_b(x,x_{ml})$, and the emission angle $\theta(x,x_{ml})$. The latter represents an irrelevant degree of freedom, which can be neglected, whereas the function $x_b(x,x_{ml})$ is exactly the linear relation anticipated in Eq. \eqref{eq:linear}, with coefficients
\begin{align}
\alpha(\delta) &=  -\frac{1}{M}\left(1+\delta\,\frac{f_{\ell}}{M}\right)\\
\beta(\delta) &=  \delta\,\frac{M}{f},
\end{align}
where $f_{\ell}$ is the focal length of the imaging lens.

\section{Conclusions}
We have proposed and implemented a new correlation imaging scheme, named light field ghost imaging, combining the usual advantage of GI (i.e., the use of a single point detector receiving light from the scene of interest and the intrinsic robustness to distortion and scattering) with the possibility of refocusing, in post-processing, different planes in the scene. Prior knowledge of the object's distance is thus unnecessary and 3D reconstruction can, in principle, be obtained. This is of the utmost importance since it enables real-time focusing, which is, in general, not possible in correlation imaging, including GI. 

This advancement is obtained by measuring on the reference beam, at the same time, the spatial and momentum distribution of light via LFI. The acquisition of complete information on the electromagnetic field allows one to calculate its structure backward till the plane of interest. Remarkably, unlike computational GI \cite{bromberg2009ghost,shapiro2008computational,sun20133d,xu2018fps}, the proposed technique offers the possibility to reconstruct three-dimensional scenes without any prior knowledge of the intensity patterns on the object; LFGI, in fact, only relies on statistical averages. Such a feature allows, on the one hand, to exploit fast and uncontrolled sources such as natural ones (see, e.g., Ref.~\cite{liu2014lensless}) and, on the other hand, to envisage proper extensions of the LFGI protocol to quantum light.

We stress that one of the difficulties in combining GI and LFI is that the field to be reconstructed is essentially a thermal incoherent field with a limited divergence and a number of transverse wave vectors. This differs from the typical direct LFI scenario, where a scattering object diffuses the light field. In LFGI, the plenoptic camera must be configured for this particular task, and the final 3D capabilities are also related to the speckle features. This dependence deserves to be investigated in depth elsewhere, and it will allow further optimization of the protocol.

Although this work represents the first proof of principle of this novel technique, we believe LFGI will allow to largely extend the range of applicability of GI in remote sensing, including the perspective of faster imaging and ranging applications. Let us notice that the principle can also be applied to the spatial characterization of detectors \cite{avella2016absolute,b,c,d}, including depth information, where the surface of the detector can play the role of the 'object', and its integrated electric output signal can be correlated with the reference pixels array. In this perspective, LFGI can be helpful for the characterization of quantum technology devices, and innovative bio-medical applications can also be devised. One example can be the investigation of the retina spatial response to light stimuli for the identification of damaged areas or specific pathologies and exploiting a source with quantum correlations; the technique could also help in investigating the vision process at single photon level \cite{Rieke,Tinsley,Field, Kelber,sim}. 

\section*{Conributions}

AA has proposed the concept of LFGI, with contribution of AM, IRB, MD and FVP. 
AP has conducted the experiment, the preliminar simulations and the experiment data analysis under the supervision of AA, AM and IRB. 
GM has conducted the formal analysis and simulations, and contributed to the interpretation of the results by developing the theoretical model under the supervision of MD and FVP.
MD and IRB provided project administration and funding acquisition.
IRB, MD, and MG (head of the INRiM Quantum Optics and Photometry sector) supervised the project. 
All authors contributed to write and agreed to the published version of the manuscript

\begin{acknowledgments}
This project has received funding from the European Defence Fund (EDF) under grant agreement EDF-2021-DIS-RDIS-ADEQUADE (n°101103417) and from INFN through the project QUISS. M.D. is supported by PNRR MUR project PE0000023 - National Quantum Science and Technology Institute. F.V.P. is supported by PNRR MUR project CN00000013 - National Centre for HPC, Big Data and Quantum Computing. Funded by the European Union. Views and opinions expressed are however those of the author(s) only and do not necessarily reflect those of the European Union or the European Commission. Neither the European Union nor the granting authority can be held responsible for them.
\end{acknowledgments}


%

\end{document}